\documentclass[12pt]{article}

\usepackage{xspace}

\def\met{$E_{\mathrm{T}}^{\mathrm{miss}}$\xspace}
\def\ptmissvec{${\vec p_{\mathrm{T}}^{\mathrm{miss}}}$\xspace}
\def\etmissvec{${\vec E_{\mathrm{T}}^{\mathrm{miss}}}$\xspace}

\usepackage[dvipdfm]{graphicx}
\usepackage{epsfig}
\usepackage{color}
\usepackage{cite}

\usepackage{hyperref}
\usepackage{authblk}
\usepackage{amssymb,amsmath}

\begin{document}

\title{Prospects on the search for invisible Higgs decays in the $ZH$ channel at the LHC and HL-LHC: \\ A Snowmass White Paper}

\author[1]{Hideki Okawa}
\author[2]{Josh Kunkle}
\author[2]{Elliot Lipeles}

\affil[1]{Brookhaven National Laboratory, Upton, NY, USA}
\affil[2]{University of Pennsylvania, Philadelphia, PA, USA}

\date{}
\maketitle

\begin{abstract}

We show prospects on a search for invisible decays of
a Higgs boson at the Large Hadron Collider (LHC) and High Luminosity
LHC (HL-LHC). 
This search is performed on a Higgs boson produced in association 
with a $Z$ boson. 
We expect that the branching ratio of 17-22\% (6-14\%) could be excluded 
at 95\% confidence level with 
300 fb$^{-1}$ (3000 fb$^{-1}$) of data at $\sqrt{s}=14$ TeV. The range
indicates different assumptions on the control of systematic uncertainties.
Interpretations with Higgs-portal dark matter models are also 
considered. 
\end{abstract}

\section{Introduction}
\label{sec:intro}

The nature of dark matter is an outstanding question in particle
physics and cosmology. One possible explanation is a weakly
interacting massive particle (WIMP) that is thermally produced in the
early universe. If such a WIMP exists and has a mass less than half
the Higgs mass, the Higgs could decay into WIMP pairs, which would
result in the Higgs boson having an invisible branching ratio
larger than the Standard Model expectation.
Limits on the Higgs to invisible branching ratio from the Large 
Hadron Collider (LHC) place constraints on
the Higgs to WIMP couplings comparable to direct detection experiments.
For low WIMP masses ($\lesssim 10$ GeV), direct detection limits are significantly weaker 
because recoil energy of a struck nucleon is lower and the corresponding signal is more 
difficult to differentiate from backgrounds. This is the region where Higgs decay 
is particularly sensitive leading to a strong complementarity with direct detection
experiments (described in Section \ref{sec:higgs_portal}).

We describe here the estimated future sensitivity of searches for Higgs decaying
invisibly using the $ZH$ channel~\cite{Godbole:2003it,Ghosh:2012ep}.  
Both ATLAS and CMS have reported
preliminary limits on the invisible branching ratio of the Higgs
using the $ZH$ channel with $Z$ decaying to electrons or muons 
and the invisible Higgs identified by a missing energy signature
\cite{zhinvATLAS,zhinvCMS}. The 95\% confidence
level (CL) limits for ATLAS and CMS are 65\% observed, 84\% expected and
75\% observed, 91\% expected, respectively. 
The vector boson fusion channel is known to have comparable 
sensitivity~\cite{Eboli:2000zeb,DiGirolamo:685420,invATLAS}, and 
CMS recently reported 95\% CL limits of 
69\% observed and 53\% expected on the invisible branching 
fraction~\cite{vbfinvCMS}. 
As a complementary approach, combined coupling measurements of the Higgs boson
allow for setting an indirect bound on the invisible and undetectable 
modes, which is 60\% at 95\% CL for ATLAS~\cite{HcouplATLAS}.

For the $ZH$ channel, the significant backgrounds for this search in
order of importance are: $ZZ \rightarrow \ell^{+}\ell^{-}\nu\bar{\nu}$, 
$WZ\rightarrow \ell\nu \ell^{+}\ell^{-}$ where one lepton is not identified, 
$W^+W^- \rightarrow \ell^{+}\nu\ell^{-}\bar{\nu}$, 
$t\bar{t}\rightarrow b\bar{b}\ell^{+}\nu\ell^{-}\bar{\nu}$ and 
$Wt \rightarrow \ell\nu b \ell\nu$
which are suppressed by a jet veto, $Z\rightarrow \ell^+\ell^-$ with false
missing energy, and $W$+jets, $t\bar{t}\rightarrow b\bar{b}\ell\nu qq$
and s/t-channel single top quark with a jet misidentified 
as a lepton. There are two main issues in scaling this result to higher
luminosities: the effect of pile-up on the $Z$ background and the
systematic uncertainty on the $ZZ$ background. It should be noted that
requirement for the angular difference between a track-based missing 
transverse momentum (\ptmissvec) and object-based missing 
transverse energy (\etmissvec; \met~for the magnitude) is not used in 
the Delphes study, which could lead to a weaker estimated limit than 
what might 
actually be attainable. On the other hand, optimistic assumptions on pileup
effects, if any, especially on the \met and jets could lead to a stronger 
estimated limit.

\section{Signal and Background Samples}
\label{sec:samp}

The signal and background simulation samples are 
generated with MADGRAPH5~\cite{madgraph5} with 
PYTHIA6~\cite{pythia} parton showering and hadronization. 
All of these samples are processed through the 
Delphes fast detector simulation~\cite{delphes1,delphes2,delphes3}. 

We used the official Snowmass background samples, and privately 
produced signal samples following the official Snowmass 
configurations for the pileup scenarios: $\langle \mu \rangle=50$ 
($\mu$: mean number of interactions per crossing)
for the LHC Phase-I with expected integrated luminosity of 
300 fb$^{-1}$ and $\langle \mu \rangle=140$ for the LHC Phase-II
also known as the High Luminosity LHC (HL-LHC) with 3000 fb$^{-1}$. 
All the processes are normalized to the 
next-to-leading order (NLO) cross sections. 
For the invisible Higgs boson signal, the 
mass of 125 GeV and the Standard 
Model cross section value of the $ZH$ production are assumed.

\section{Object and Event Selection}
\label{sec:evsel}

We use the reconstructed objects from the Delphes 
detector simulation, which include contributions from the pileup. 
Events are selected to have two charged leptons 
(electrons or muons) and large \met. 
While the \met requirement suppresses the Drell-Yan
background, an additional jet-veto is used to suppress the top backgrounds.

\begin{figure}[t]
  \begin{center}     
  \includegraphics[width=0.49\textwidth]{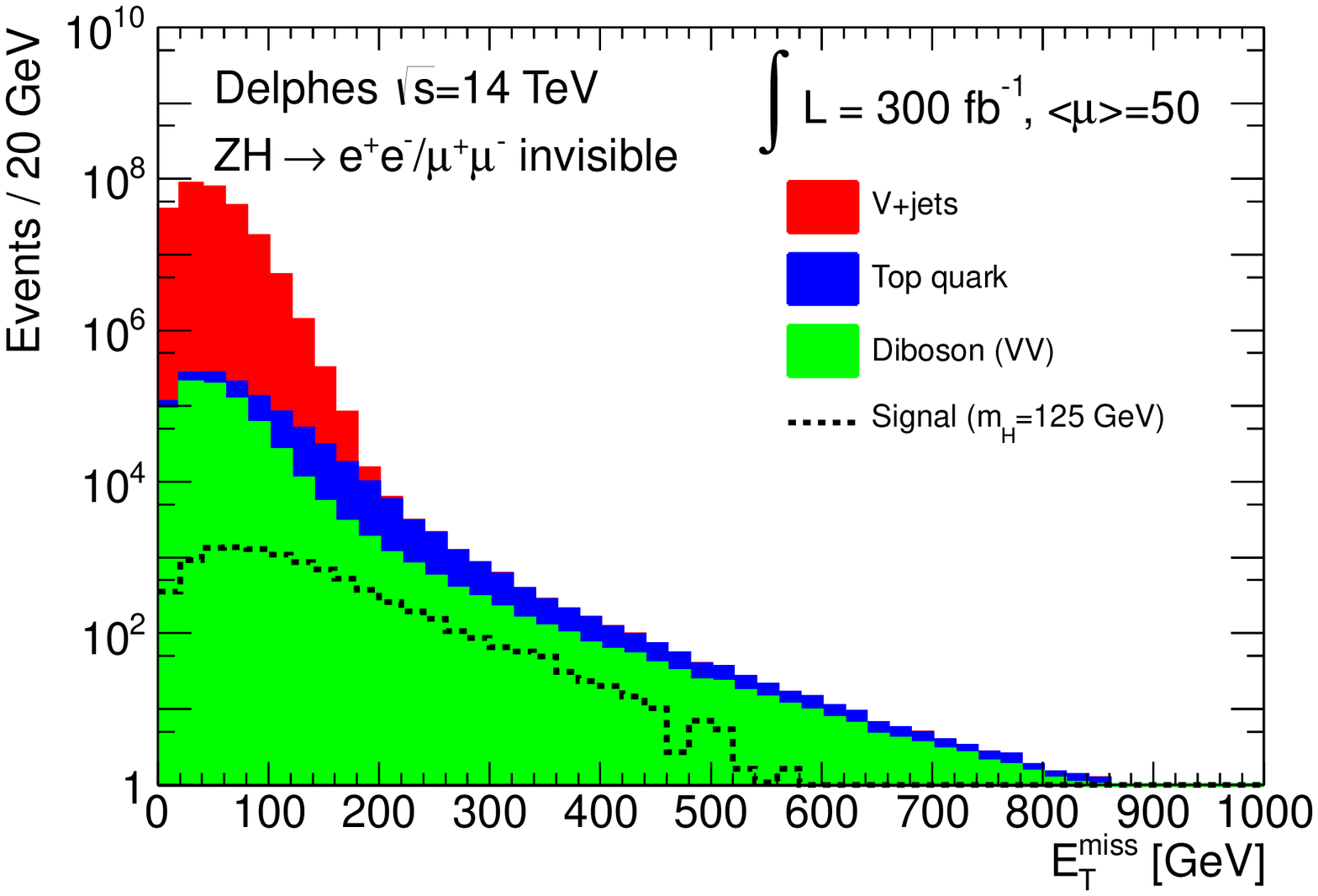}
  \includegraphics[width=0.49\textwidth]{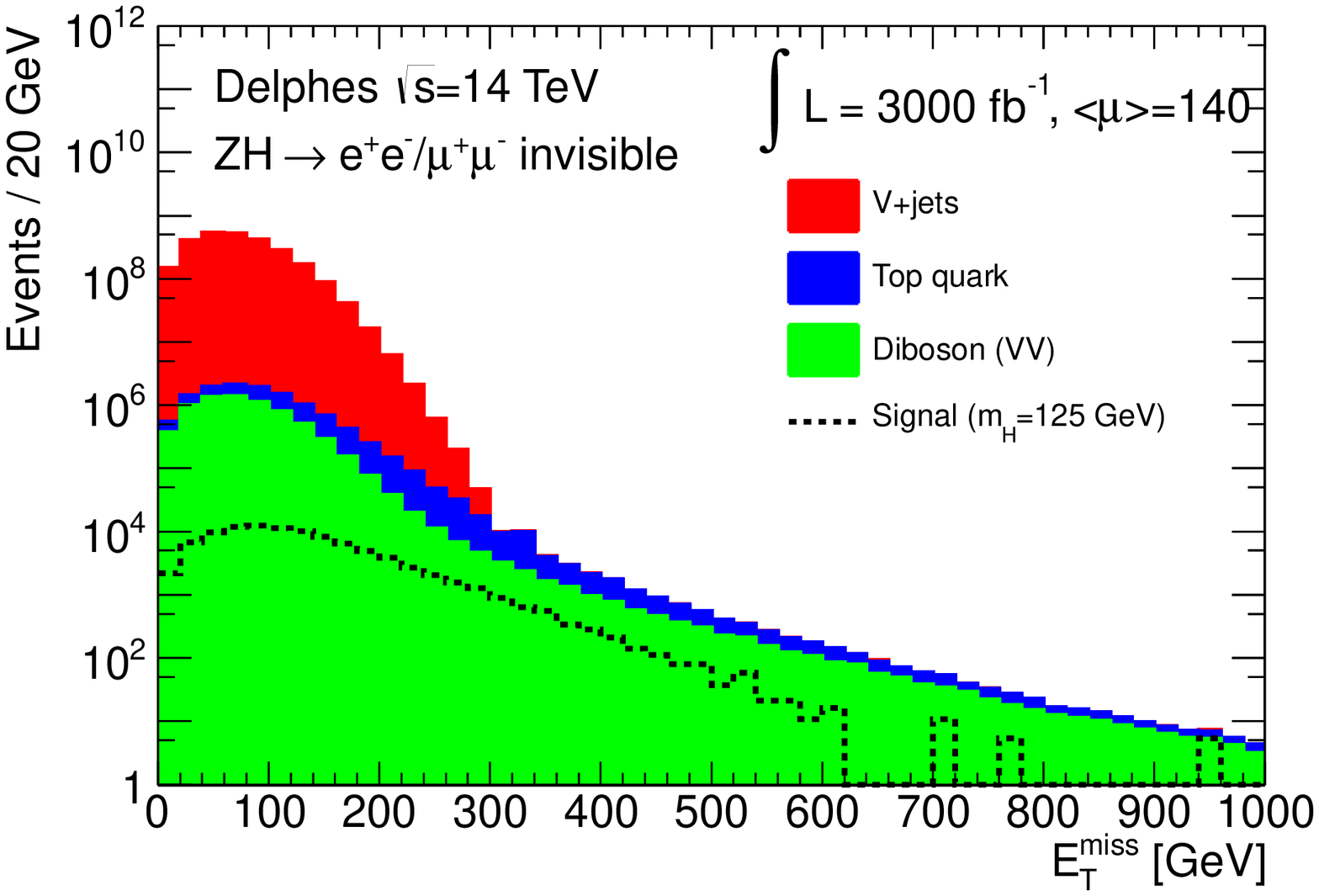}
  \end{center}     
\caption{ \label{fig:met_afterMZ} \met~distributions after the dilepton mass
  requirement for the 14 TeV LHC and HL-LHC scenarios. The stacked histograms represent the background predictions from the Delphes samples. The Delphes samples are inclusively generated for various bosons, and ``V'' stands for $W$, $Z$, $\gamma$, and $H$. The signal hypothesis is shown by a dotted line and assumes the SM $ZH$ production rate for a $m_H = 125$ GeV Higgs boson with a 100\% invisible branching ratio.}
\end{figure}

As we consider events with a $Z$ boson and large \met, we
require two oppositely-charged electrons or muons with an invariant mass
between 76 to 106 GeV.
Electrons (muons) are required to have $p_T > $ 20 GeV and $|\eta| < $
2.47 (2.5). In order to reduce the $WZ\rightarrow \ell\nu \ell^+\ell^-$
background, events with additional lower threshold electrons and muons
($p_T > $ 7 GeV) are vetoed.
Figure~\ref{fig:met_afterMZ} shows the \met
distributions for the events after requiring a dilepton in the $Z$
mass window.

\begin{table} [t]
  \begin {center}
\begin{tabular}{|c|c|c|}
\hline
   Cut variables &  Thresholds (LHC, $\langle \mu \rangle$=50) & Thresholds (HL-LHC, $\langle \mu \rangle$=140)   \\  \hline
   \met & $>$ 150 GeV  & $>$ 170 GeV \\ \hline
   d$\phi(\ell,\ell)$ & $<$ 1.4 & $<$ 1.1 \\ \hline
   d$\phi(\vec{p}_T^{\ell,\ell},\vec{E}_{\mathrm{T}}^{\mathrm{miss}})$ & $>$ 2.6 & $>$ 2.6 \\ \hline
   $|$\met - $p_T^{\ell,\ell}|/p_T^{\ell,\ell}$ & $<$ 0.3 & $<$ 0.4 \\ \hline
   Jet veto & $>$ 45 GeV & $>$ 60 GeV \\ \hline
\end{tabular}
\caption{\label{tab:evsel} Event selection optimized for the LHC and HL-LHC scenarios at 14 TeV.} 
\end{center}
\end{table}

\begin{figure}[t]
  \begin{center}     
  \includegraphics[width=0.49\textwidth]{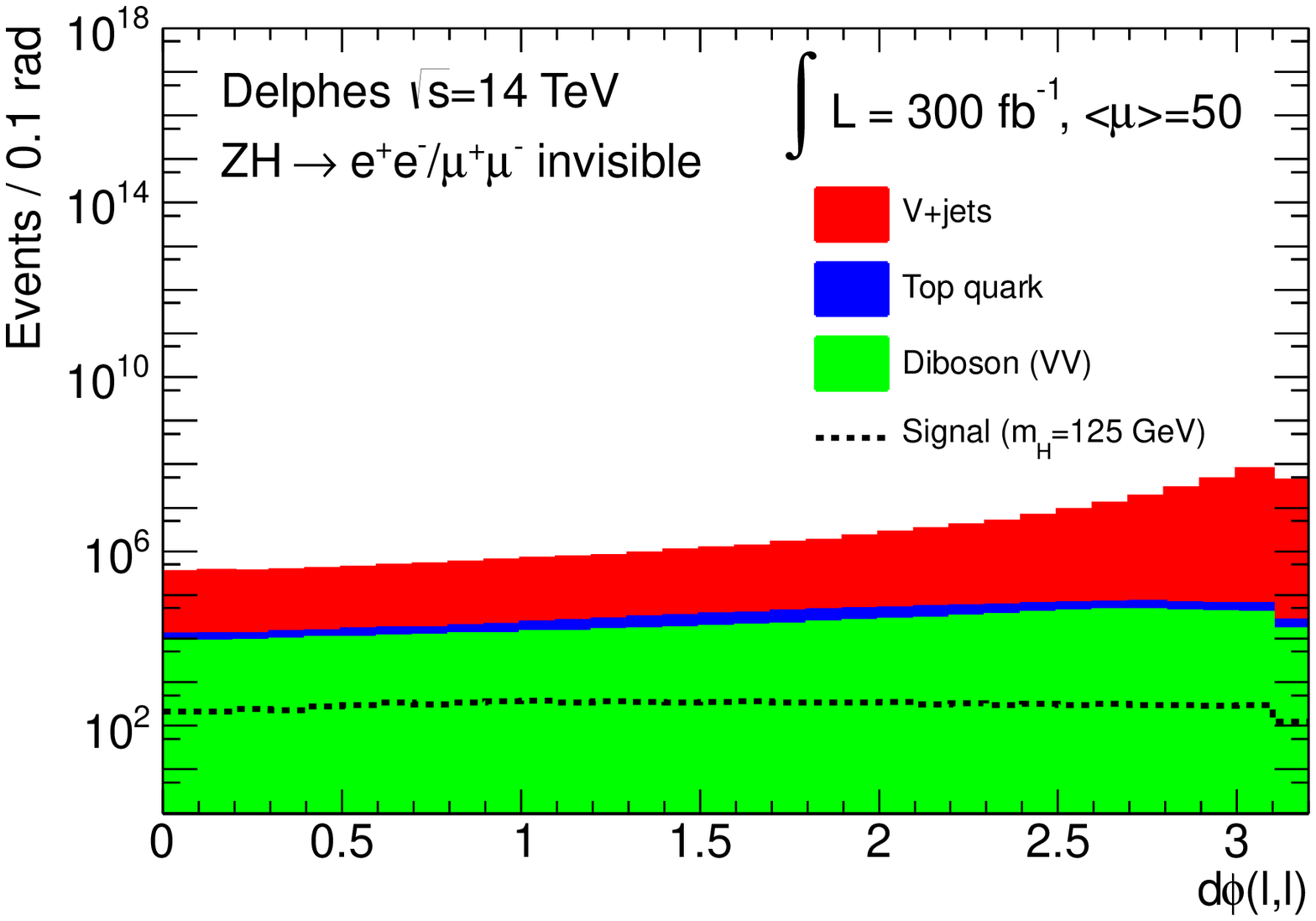}
  \includegraphics[width=0.49\textwidth]{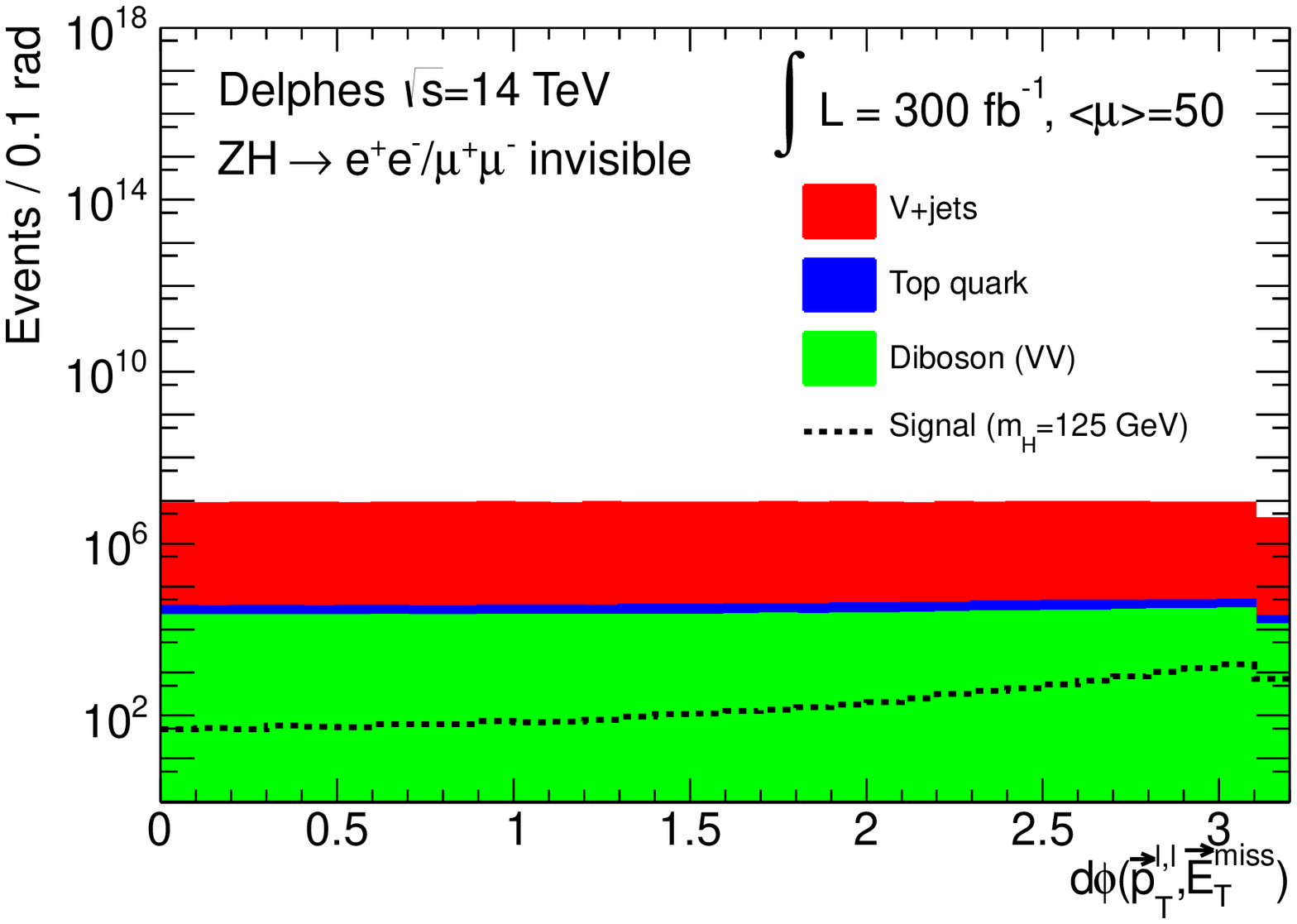} \\
  \includegraphics[width=0.49\textwidth]{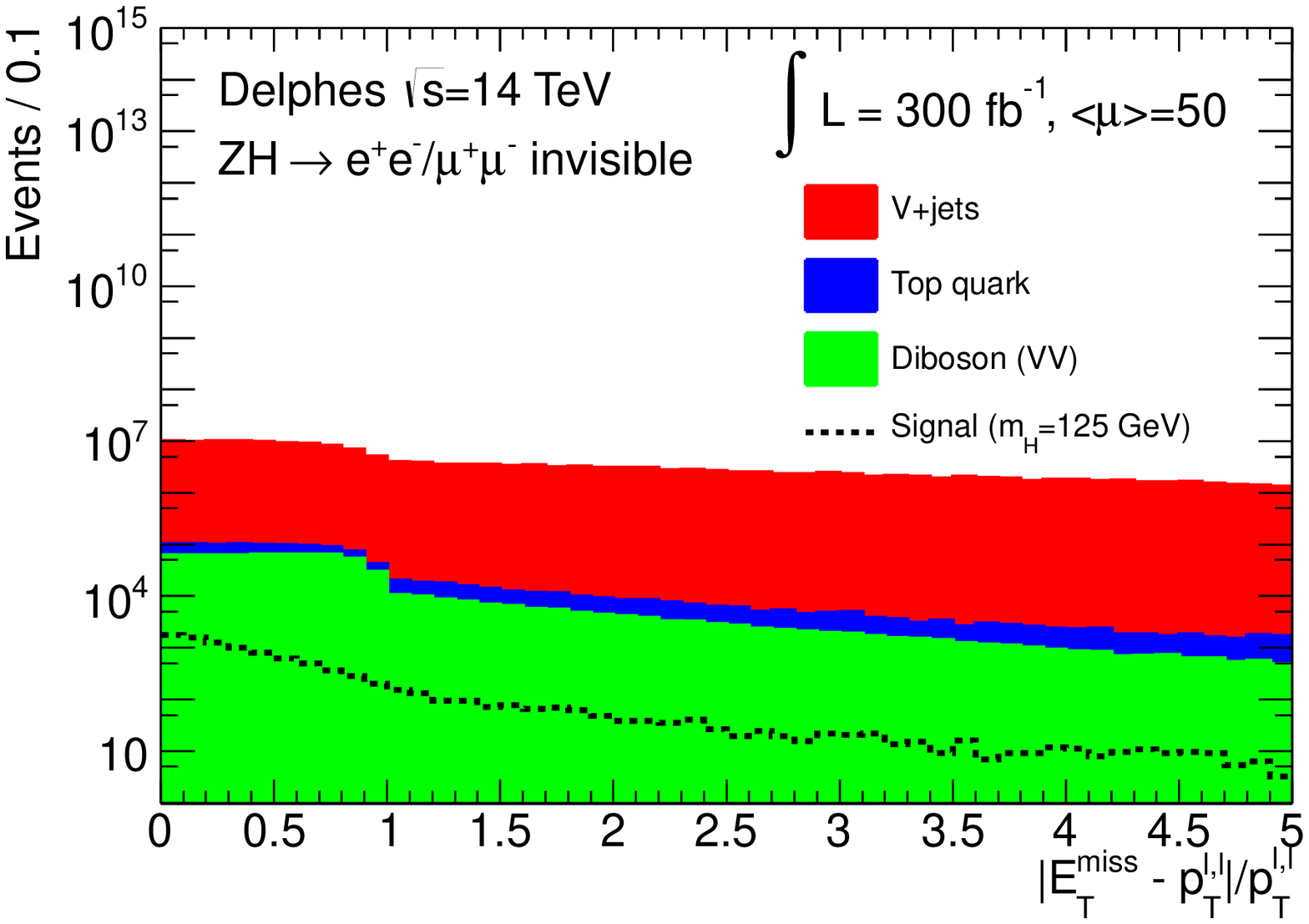}
  \includegraphics[width=0.49\textwidth]{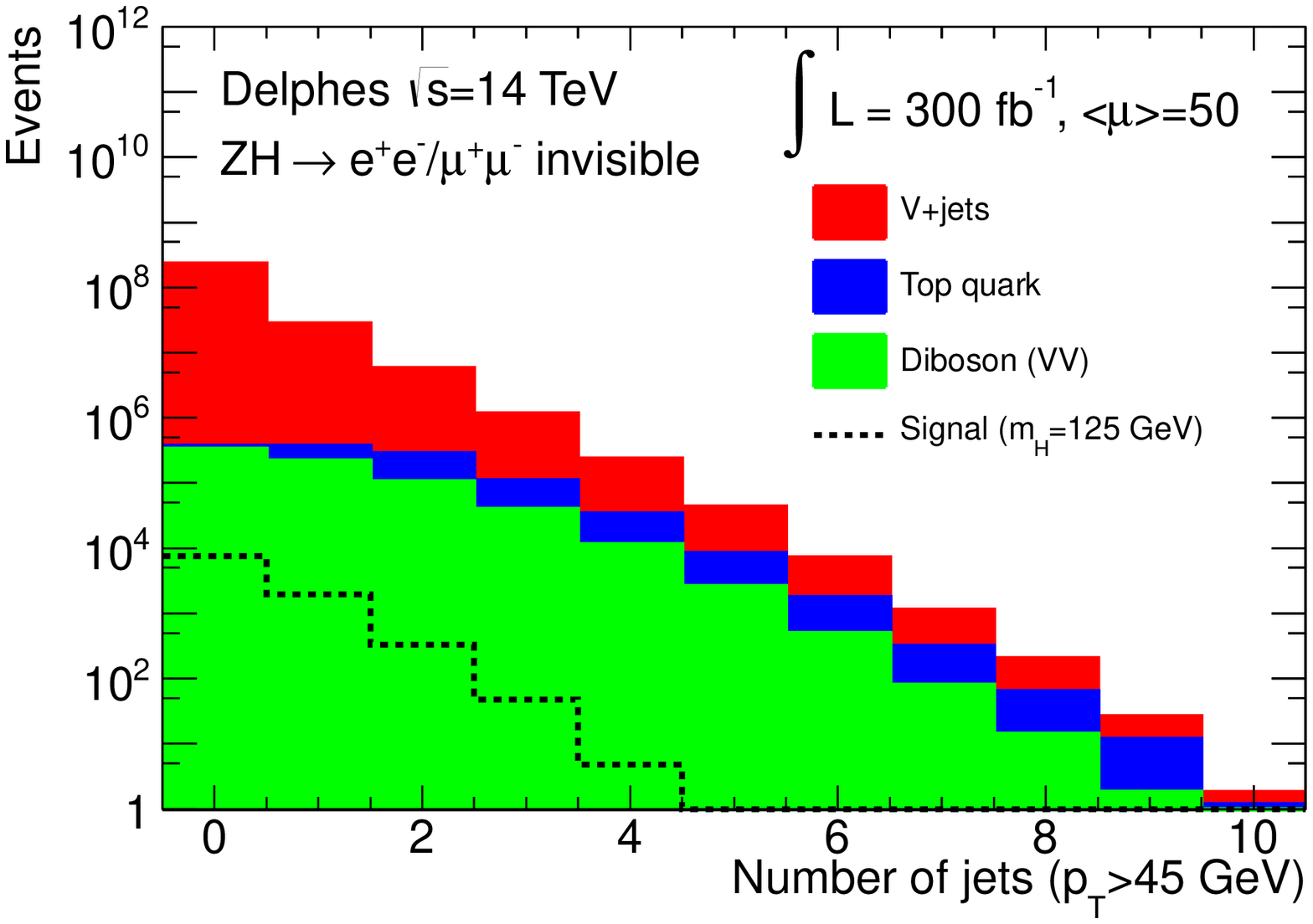}
  \end{center}     
  \caption{\label{fig:cutvar_afterMZ} Distributions of kinematic variables after the dilepton mass
  requirement for the LHC scenario. The stacked histograms represent the background predictions from the Delphes samples. The Delphes samples are inclusively generated for various bosons, and ``V'' stands for $W$, $Z$, $\gamma$, and $H$. The signal hypothesis is shown by a dotted line and assumes the SM $ZH$ production rate for a $m_H = 125$ GeV Higgs boson with a 100\% invisible branching ratio.}
\end{figure}

Our analysis is based on the ATLAS event selection~\cite{zhinvATLAS}
with the following modifications. 
\begin{itemize}
\item The \met~is required to be larger 
than 150 GeV for 300 fb$^{-1}$ and 170 GeV for 3000 fb$^{-1}$.
\item The \ptmissvec~is not considered. 
\item The $p_T$ threshold for the jet veto is raised to 45 GeV for 300 fb$^{-1}$ 
and 60 GeV for 3000 fb$^{-1}$, as the jet reconstruction in the Delphes
simulation does not correct for the pileup. 
\item Some of the angular cuts with the \etmissvec are relaxed, because the 
correlation of the \etmissvec with the dilepton system is degraded due to the pileup. 
\end{itemize}
The thresholds are chosen to increase the signal sensitivity. 
The event selection used in this analysis is summarized in Table~\ref{tab:evsel}.
Figure~\ref{fig:cutvar_afterMZ} shows the distributions of the kinematic variables 
used in the event selection
for the events after requiring a dilepton in the $Z$ mass window. 

\section{Results}
\label{sec:results}

Table~\ref{tab:yields} shows the expected background and signal yields 
for the two luminosity scenarios. Only the statistical uncertainty from 
the Delphes samples is shown in the table. 
The $Z$ and $W$ backgrounds do not 
remain after applying the event selection for the LHC scenario. 
Very few top events remain
after the jet veto cut, and thus lead to a large statistical uncertainty. 

\begin{table} [t]
  \begin {center}
\begin{tabular}{|c|c|c|}
\hline
   Expected yields &  LHC (300 fb$^{-1}$) & HL-LHC (3000 fb$^{-1}$)  \\  \hline
   Dibosons ($ZZ$, $WZ$, $WW$, etc.) & $1754 \pm 29$ & $12009 \pm 203$ \\ \hline
   $t\bar{t}$, single top & $2.4 \pm 1.4$ & $550 \pm 530$ \\ \hline
   Single boson ($Z/W$ + jets, etc.) & -- & $1199 \pm 599$ \\ \hline\hline
   Signal (125 GeV, BR($H\rightarrow$ inv.)=20\%) & $217 \pm 5$ & $1517 \pm 40$ \\\hline
\end{tabular}
\caption{\label{tab:yields} Expected background and signal yields for the LHC and HL-LHC scenarios.
The statistical uncertainty from the Delphes samples is shown.} 
\end{center}
\end{table}

\subsection{Systematics}
\label{sec:sys}

We consider two scenarios for the size of systematics.

For the conservative scenario, experimental uncertainty of 5\%, 
theoretical uncertainty of 5\%, and jet veto systematics of 
6\% are assumed for the diboson backgrounds.
 
For the realistic case, the uncertainty is expected to become
smaller using data-driven methods, making use of the large data 
statistics. From the expected yields of 
the $ZZ\rightarrow 4\ell$, the $ZZ$ background would be estimated 
within 
6\% for 300 fb$^{-1}$ and 2\% for 3000 fb$^{-1}$. We adopt this 
uncertainty for the diboson backgrounds. 

For both scenarios, the top quark background is estimated
to have the overall uncertainty of 9\% for 300 fb$^{-1}$ 
and 3\% for 3000 fb$^{-1}$, 
extrapolating the expected yields in the e$\mu$ control region from
Ref.~\cite{zhinvATLAS}.  

The $Z$ background is assumed to have the uncertainty of 10\%, but 
this background is expected to be suppressed significantly by the 
d$\phi$(\etmissvec, \ptmissvec) selection, which is not applied in this 
paper. 

For the signals, the experimental uncertainty of 5\%, theoretical 
uncertainty of 5\%, and jet veto systematics of 6\% are 
considered for all cases. 

During the limit setting, the correlation 
between the signals and the diboson backgrounds is taken into 
account for the jet veto systematics.

\subsection{Sensitivity}
\label{sec:reach}

We calculated the limits with the $CL_s$ modified frequentist 
formalism~\cite{CLspaper} 
using a maximum likelihood fit using the \met distributions with a 
profile likelihood test statistics~\cite{asymptotic}.  

Table~\ref{tab:limits} shows the expected limits with various size of 
systematic uncertainty on the background and signal. 
For the HL-LHC scenario, the $Z$ background is concentrated at the
lowest \met bin, thus the signal sensitivity still remains to be high. 
The 90\% CL limits
are also shown to be compared with direct detection dark matter 
experiments in Section~\ref{sec:higgs_portal}. 
The dominant
systematics is the $ZZ$ normalization. The current LHC results 
\cite{zhinvATLAS,zhinvCMS} quote uncertainties between 7\% and 11\%.
The Higgs to $WW$ measurements are able to use control regions to normalize
similar backgrounds to better than 2\% \cite{ATLAS_HWW_Moriond2013}. 

\begin{table} [t]
  \begin {center}
\begin{tabular}{|c|c|c|}
\hline
   BR($H\rightarrow$inv.) limits at 95\% (90\%) CL &  LHC (300 fb$^{-1}$) & HL-LHC (3000 fb$^{-1}$)  \\  \hline
   No systematics & 7.5\% (6.2\%)  & 2.9\% (2.5\%) \\ \hline
   Realistic scenario & 17\% (14\%)  &  6.2\% (5.2\%) \\ \hline
   Conservative scenario &  22\% (19\%) &  14\% (11\%) \\ \hline
\end{tabular}
\caption{\label{tab:limits} Expected limits with 95\% (90\%) CL on the invisible branching ratio of the Higgs boson are shown for the LHC and HL-LHC scenarios. The Standard Model cross section for the $ZH$ production is assumed.} 
\end{center}
\end{table}

\subsection{Interpretation with Higgs-Portal Models}
\label{sec:higgs_portal}

A possible interpretation of the invisible decay of the Higgs boson 
is in the context of dark matter particles coupling to the 
Higgs boson. Such dark matter models are called the Higgs-portal 
models~\cite{higgs_portal,higgs_portal3,higgs_portal1,higgs_portal2}. 

The model considered here introduces dark matter as a 
single new particle that couples only with the Higgs boson.
The strength of the interaction between the dark matter and the Higgs boson is 
given by the coupling constant, $\lambda_{h\chi\chi}$.  
Within the Higgs-portal models, limits from the invisible
branching ratio of the Higgs can be compared to
limits from dark matter direct detection experiments. 
This is possible because the scattering process to which
direct detection experiments are sensitive is related to the
decay process used here by the coupling constant.
The relationship between the decay width, the coupling constant, and 
the dark matter-nucleon scattering cross section depend on the 
spin of the dark matter particle~\cite{higgs_portal3,higgs_portal1,higgs_portal2,DM_spin_xsec}. 
We consider three spin
scenarios: a scalar, vector, or majorana-fermion. 

\begin{figure}[p]
  \begin{center}     
  \includegraphics[width=0.7\textwidth]{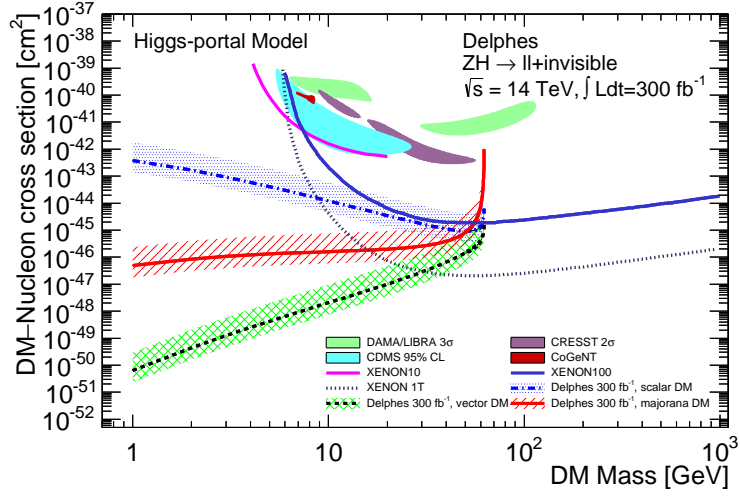}
  \includegraphics[width=0.7\textwidth]{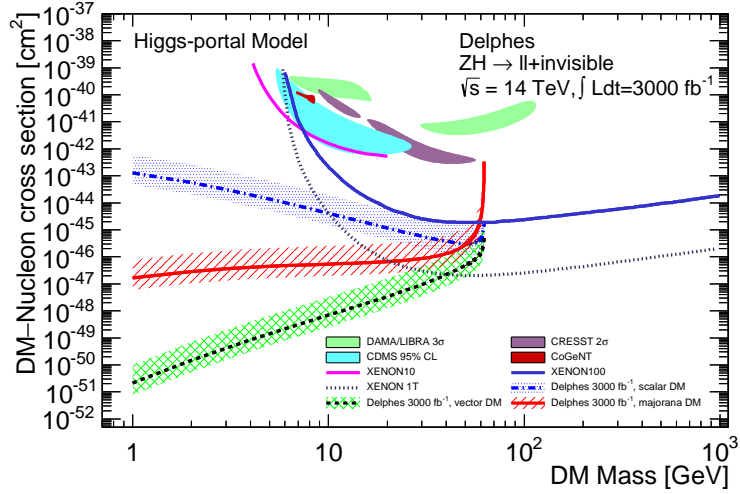}
  \end{center}     
\caption{ \label{fig:higgsportal_xsec} 90\% C.L. upper limits on the dark matter-nucleon scattering cross section in Higgs-portal scenarios, extracted from the expected Higgs to invisible branching fraction limit and from direct-search experiments. The results are shown for three model variants in which the DM candidate is a scalar, vector or fermion particle. The hashed areas correspond to the uncertainty of the nucleon form factor.}
\end{figure}

\begin{figure}[p]
  \begin{center}     
  \includegraphics[width=0.7\textwidth]{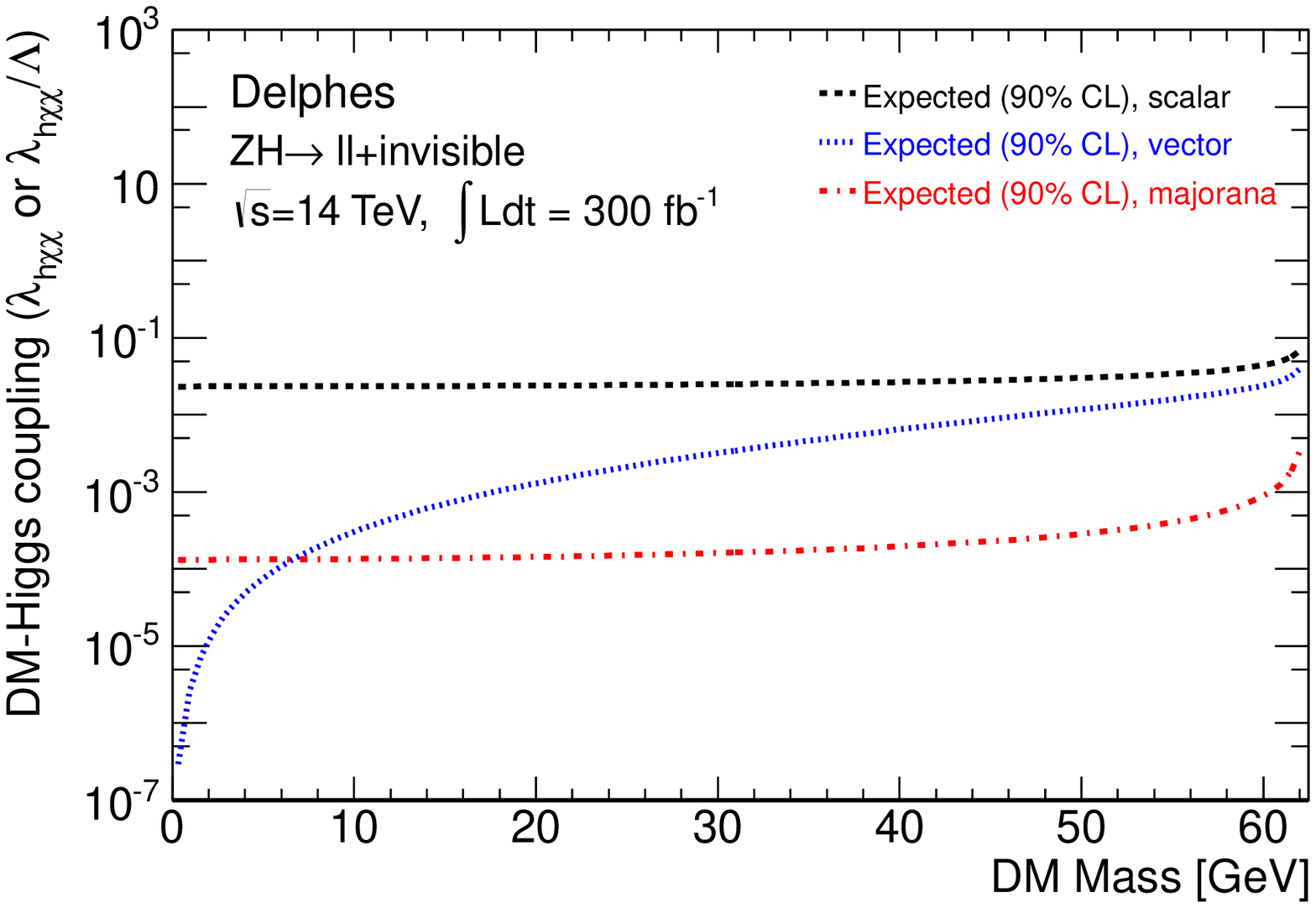} \\
  \includegraphics[width=0.7\textwidth]{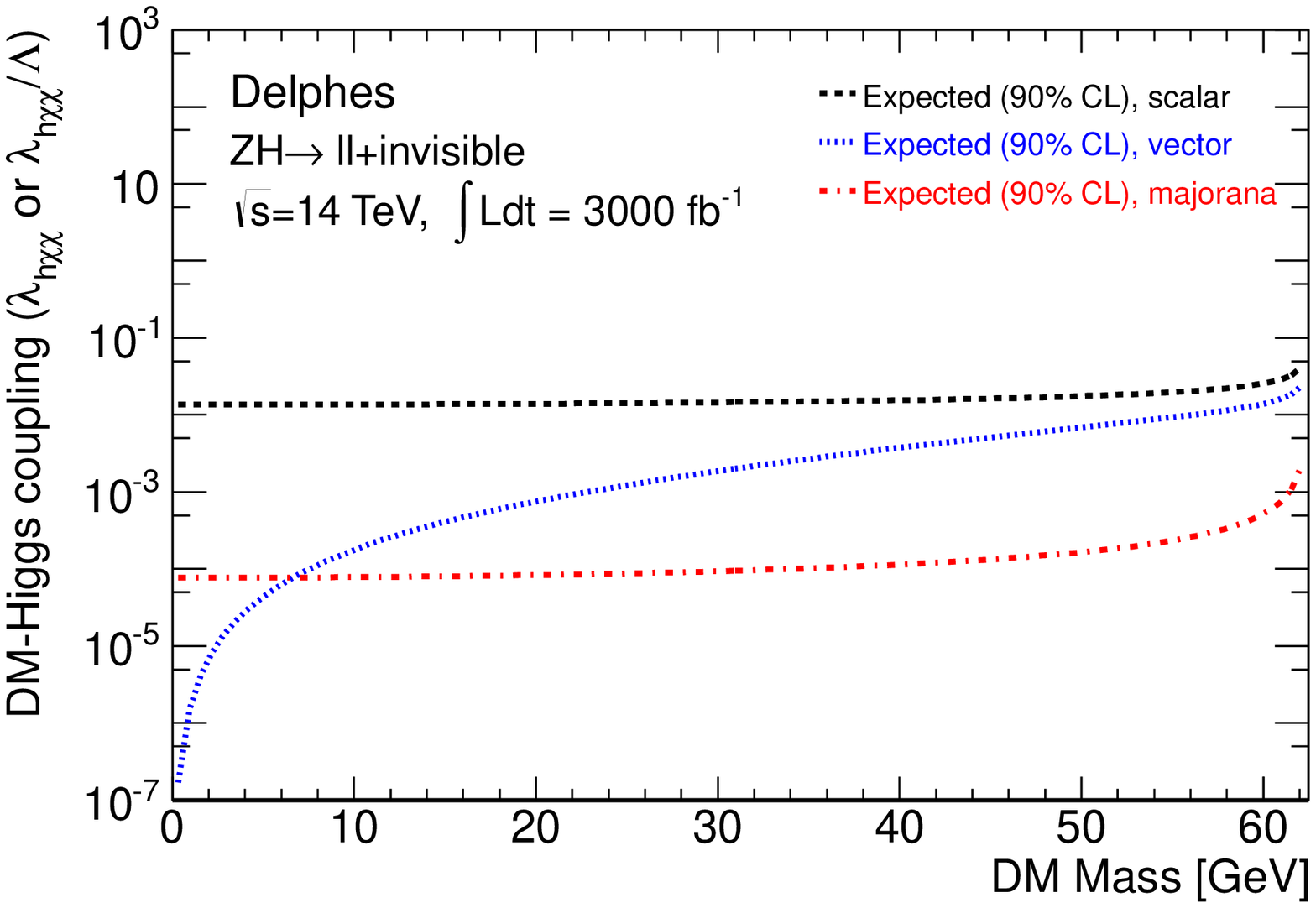} 
  \end{center}     
\caption{ \label{fig:higgsportal_coupl} 90\% CL limits on the Higgs-dark matter couplings in Higgs-portal scenarios, extracted from the expected Higgs to invisible branching fraction limit. The results are shown for three model variants in which the dark matter candidate is a scalar, vector or fermion particle.}
\end{figure}

Figure~\ref{fig:higgsportal_xsec} shows the upper limits on the 
dark matter-nucleon scattering cross section. 
The expected limits from the ``realistic scenario'' in 
Table~\ref{tab:limits} are used for the interpretation. 
In expressing the invisible branching fraction limits in 
terms of the dark matter-nucleon scattering cross section,
the nucleon form factor, $f_N$, must be included to
parametrize the coupling between the Higgs boson and
the nucleon.
The nucleon form factor is taken as 
$0.326^{+ 0.303}_{- 0.066}$~\cite{higgs_portal1}. 
Its uncertainty is expressed as systematics bands in the figure. 
To be consistent 
with the direct detection 
experiments~\cite{xenon, xenon100, xenon1T, cresst, dama, cogent, cogent2, cdms},
we use 90\% CL limits to map the branching ratio bounds. 
Figure~\ref{fig:higgsportal_coupl} shows the upper limits on the 
Higgs-dark matter couplings. 

In the context of the Higgs-portal models, the LHC has an 
outstanding sensitivity for the mass region of the dark matter 
below half the Higgs mass. 

\section{Conclusions}
\label{sec:concl}

We showed prospects on a direct search for invisible decays of
a Higgs boson at the LHC and HL-LHC. 
This search is performed on a Higgs boson produced in association 
with a $Z$ boson. 
We expect that the branching ratio of 17-22\% (6-14\%) could be excluded 
at 95\% confidence level with 
300 fb$^{-1}$ (3000 fb$^{-1}$) of data at $\sqrt{s}=14$ TeV. The range
indicates different assumptions on the control of systematic uncertainties.
We interpret the results in the context of Higgs-portal
models, which shows a strong complementarity between invisible
Higgs decay searches and direct dark matter searches.

\section*{Acknowledgments}

JK acknowledges Yann Mambrini, Joachim Kopp, and Gary Steigman for useful discussions regarding the dark matter interpretations. HO thanks Hong Ma and Marc-Andre Pleier for fruitful conversations. 
This work was supported by the US Department of Energy under 
Contract No. DE-AC02-98CH10886 and DE-SC0007901.

\bibliographystyle{whitepaperbib}
\bibliography{zhinv_whitepaper}

\end{document}